\documentclass[reprint,aps,pra,amsmath,amssymb,amsfonts,showkeys,showpacs,keywords]{revtex4-1}
\usepackage{graphicx,dcolumn,bm,color,hyperref,microtype,multirow,subfigure}

\begin{document}
\newcommand{\EHF}{E^{\rm HF}}
\newcommand{\Ec}{E^{\rm corr}}
\newcommand{\Eh}{E_{\rm h}}
\newcommand{\rij}{\hat{r}}
\newcommand{\mc}{\multicolumn}
\newcommand{\mr}{\multirow}
\newcommand{\alert}[1]{\textcolor{black}{#1}}

\title{Understanding excitons using spherical geometry}

\author{Pierre-Fran\c{c}ois Loos}
\email{loos@rsc.anu.edu.au}
\affiliation{Research School of Chemistry, Australian National University, Canberra ACT 0200, Australia}
 
\date{\today}
\keywords{Exciton; Frenkel exciton; Wannier-Mott exciton; Exact solution; Spherical geometry}
\pacs{71.35.-y, 71.35.Aa}

\begin{abstract}
Using the spherical geometry, we introduce a novel model to study excitons confined in a three-dimensional space, which offers unparalleled mathematical simplicity while retaining much of the key physics. This new model consists of an exciton trapped on the 3-sphere (i.e. the surface of a four-dimensional ball), and provides a unified treatment of Frenkel and Wannier-Mott excitons. Moreover, we show that one can determine, for particular values of the dielectric constant $\epsilon$, the closed-form expression of the exact wave function. We use the exact wave function of the lowest bound state for $\epsilon=2$ to introduce an intermediate regime which gives satisfactory agreement with \alert{the} exact results for a wide range of $\epsilon$ values.
\end{abstract}

\maketitle

{\em Excitons.---}An exciton (X) is a quasiparticle created by the association of an electron (e) and an electron hole (h) attracted to each other by the Coulomb force \cite{Excitons, MolExcitons}. The electron and hole may have either parallel or anti-parallel spins, and form an electrically neutral quasiparticle able to transport energy without carrying net electric charge. The concept of \alert{an} exciton was first proposed by Frenkel \cite{Frenkel31} in 1931 to described excitations in insulators. In semiconductors, a hole is usually created when a photon is absorbed, and excites an electron from the valence to the conduction band, yielding a positively-charged hole. In such materials, excitons play a key role in optical properties \cite{BasuBook}.

\begin{figure}
	\includegraphics[width=0.3\textwidth]{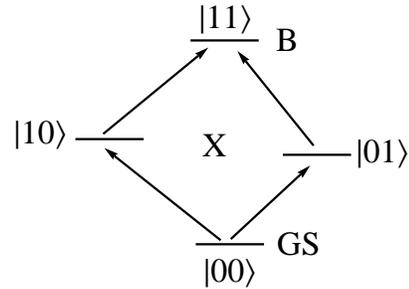}
	\caption{Four-level excitonic system as a prototype of a quantum gate.}
	\label{fig:fig1}
\end{figure}

These systems are of particular interest in quantum information and computation to construct coherent combinations of quantum states \cite{NeilsonBook}. Following DiVincenzo's theory \cite{DiVincenzo95}, quantum gates operating on just two qubits at a time are sufficient to construct a general quantum circuit. The basic quantum operations can be performed on a sequence of pairs of physically distinguishable quantum bits and, therefore, can be illustrated by a simple four-level system, as shown in Fig. \ref{fig:fig1}. In optically driven systems, direct excitation \alert{to} the upper $|11\rangle$ level (lowest biexciton\footnote{a biexciton (B) corresponds to the association of two excitons.} state) from the ground state (GS) $|00\rangle$ is usually forbidden and the most efficient alternative is to use two distinguishable excitonic states with orthogonal polarizations ($|01\rangle$ and $|10\rangle$) as intermediate states. However, in atomic systems, excitation \alert{to} the upper level does not ensure quantum coherence between $|00\rangle$ and $|11\rangle$ leading to errors in the quantum logic device \cite{Li03}.

There are two main kinds of excitons: Frenkel excitons \cite{Frenkel31} (sometimes called molecular excitons \cite{MolExcitons}) are found in materials where the dielectric constant is generally small, and are characterized by compact, localized wave functions. Wannier-Mott excitons \cite{Wannier37, Mott38} are found in semiconductors with a large dielectric constant, and have large, delocalized wave functions.

{\em The model.---}In 1983, Laughlin \cite{Laughlin83} proposed an accurate trial wave function in order to explain and predict the fractional quantum Hall effect (FQHE), and eventually received the Nobel prize in physics in 1998 (jointly with Horst L. St\"ormer and Daniel C. Tsui) for the discovery of this new form of quantum fluid with fractionally charged excitations. A few months after the publication of LaughlinÕs paper, Haldane \cite{Haldane83} introduced the spherical geometry for the study of the FQHE, wherein the two-dimensional sheet containing electrons is wrapped around the surface of a 2-sphere, and a perpendicular (radial) magnetic field is generated by placing a Dirac magnetic monopole at the centre of the 2-sphere. This geometry has played an important role in testing various theoretical conjectures. The main reason for the popularity of this compact geometry is that it does not have edges, which makes it suitable for an investigation of bulk properties. The spherical geometry has been instrumental in establishing the validity of the FQHE theory, and provides the cleanest proof for many properties \alert{\cite{JainBook}}.

Following Haldane's footsteps, we introduce a simple model using the spherical geometry to study excitons confined in a three-dimensional space for any value of the dielectric constant. It yields a unified treatment of Frenkel and Wannier-Mott excitons \cite{Egri79}, and we will show that one can determine, for a particular value of the dielectric constant, the closed-form expression of the exact wave function associated with the lowest bound state (i.e. associated with a negative energy). 

Our model consists of an exciton trapped on the 3-sphere (i.e. the surface of a four-dimensional ball). Excitons on the surface of a 2-sphere have been previously studied theoretically \cite{Kayanuma92, Pedersen10} and experimentally \cite{Kanemitsu93, Kanemitsu94}. However, our model has the advantage of having the same dimensionality as real three-dimensional solid-state or molecular systems. Moreover, previous studies on two-electron \cite{QuasiExact09, EcLimit09, Excited10, EcProof10} and many-electron systems \cite{Glomium11} have shown many similarities between real and spherically-confined  systems. 

In Ref.~\onlinecite{Pedersen10}, Pedersen reports exact solutions for the unbound states (i.e. associated with a positive energy) \alert{of} an exciton on the surface of a 2-sphere based on the recursive approach developed in Ref.~\onlinecite{QuasiExact09}. However, excitons on a three-dimensional spherical surface have not been considered before, and this Letter presents the first study of exact solutions and asymptotic regimes of excitons in a \alert{spherical} three-dimensional space. To our best knowledge, this is also the first study reporting an exact closed-form solution associated with the lowest bound state of an exciton.

{\em Schr\"odinger equation.---}Let us consider an exciton created on the surface of a 3-sphere of radius $R$. The coordinates of a particle on a 3-sphere are given by the set of hyperspherical angles $\Omega = (\theta,\phi,\chi)$. In atomic units ($\hbar=e=1$), the Schr\"odinger equation of the system is
\begin{equation}
	\label{H}
	\left(\frac{ \nabla_{\text{e}}^2}{2\,m_{\text{e}}R^2} + \frac{\nabla_{\text{h}}^2}{2\,m_{\text{h}}R^2} - \frac{1}{u}\right)  \Psi(\Omega_\text{e},\Omega_\text{h})= \mathcal{E}\, \Psi(\Omega_\text{e},\Omega_\text{h}),
\end{equation}
where 
\begin{multline}
	\nabla^2 =\frac{1}{\sin^2 \theta}\left[\frac{\partial}{\partial\theta}\left(\sin^2\theta\frac{\partial}{\partial\theta}\right)\right.
	\\
	+ \left.\frac{1}{\sin\phi}\frac{\partial}{\partial\phi}\left(\sin\phi\frac{\partial}{\partial\phi}\right) + \frac{1}{\sin^2 \phi} \frac{\partial^2}{\partial\chi^2}\right]
\end{multline}
is the Laplace operator in hyperspherical coordinates \cite{Louck60}, $m_\text{e}$ and $m_\text{h}$ are the masses of the electron and the hole, and 
\begin{equation}
	\alert{
	u^{-1} = \left| \bm{r}_1 - \bm{r}_2 \right|^{-1} = (R \sqrt{2-2\cos\omega})^{-1}
	}
\end{equation}
is the Coulomb interaction between the two particles, where the cosine of the interparticle angle is
\begin{equation}
	\alert{
\begin{split}
	\cos \omega  
	& = \frac{r_1^2 + r_2^2 - u^2}{2\,r_1\,r_2}  
	\\
	& = \cos\theta_\text{e} \cos\theta_\text{h} + \sin\theta_\text{e} \sin\theta_\text{h} \cos\phi_\text{e} \cos\phi_\text{h} 
	\\
	& \quad +  \sin\theta_\text{e} \sin\theta_\text{h} \sin\phi_\text{e} \sin\phi_\text{h} \cos\left(\chi_\text{e}-\chi_\text{h}\right).
\end{split}
	}
\end{equation}

Here, we consider exciton states with zero angular momentum and in which the two particles have opposite spin (singlet states). The present study can be easily generalized to higher angular momentum states for both the singlet and triplet manifolds \cite{Excited10}.

For zero angular momentum states, $\Psi$ depends only on the relative coordinate $\omega$ or $u$. An analysis of \eqref{H} reveals that $\Psi$, like the para-positronium wave function, possesses an ``anti-Kato'' behavior \cite{Kato57} 
\begin{equation}
\label{cusp}
	\left.\frac{\Psi^{\prime}(u)}{\Psi(u)}\right|_{u=0} = -\frac{1}{2},
\end{equation}
which shows that $\Psi$ must behave as $\Psi(u) = 1-u/2+ O(u^2)$ for small interparticle distance.

After a suitable scaling of energy ($E\leftarrow \mu\,\mathcal{E}\,R^2$), the Schr\"odinger equation \eqref{H} is  
\begin{equation}
\label{H-omega}
	\Psi^{\prime\prime}(\omega)
	+2 \cot\omega\,\Psi^{\prime}(\omega)
	+ \left( \frac{1}{\epsilon\sqrt{2-2\cos\omega}}+E\right)\,\Psi(\omega) = 0,
\end{equation}
where $\epsilon = 1/(\mu\,R)$ can be regarded as the relative dielectric constant of the system, and $\mu = 2\,m_\text{e}\,m_\text{h}/(m_\text{e} + m_\text{h})$ is the reduced mass of the exciton \cite{Mattis84}.

\begin{figure}
	\includegraphics[width=0.4\textwidth]{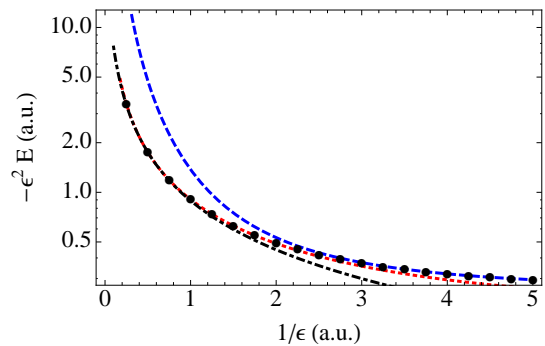}
	\caption{Energy for the lowest bound state of an exciton on a 3-sphere as a function of $\epsilon^{-1}$ in the Frenkel (dashed blue), intermediate (dot-dashed black), and Wannier-Mott (dotted red) regimes. \alert{Exact results are also reported (circles).}}
	\label{fig:fig2}
\end{figure}

\begin{table*}
\caption{
\label{tab:values}
\alert{First-order Frenkel, intermediate, Wannier-Mott and exact energies for various $\epsilon$ values. The deviation with respect to the exact result is given in parenthesis.
}}
\begin{ruledtabular}
\begin{tabular}{cccccc}
$\epsilon$	&	$E_\text{F}^{(1,1)}$	&	$E_\text{int}^{(1,1)}$		&	$E_\text{WN}^{(1,1)}$	&	$E^{(1)}$\\
\hline
 0.25 	& -5.12500 (0.00128) & -3.60055 (1.52573) & -3.39531 (1.73097) & -5.12628 \\
 1/3 		& -3.37500 (-0.02363) & -2.69682 (0.65455) & -2.54648 (0.80489) & -3.35137 \\
 0.5		& -2.12500 (-0.13978) & -1.79309 (0.19213) & -1.69765 (0.28757) & -1.98522 \\
 1.0		& -1.37500 (-0.46828) & -0.88936 (0.01736) & -0.84883 (0.05790) & -0.90672 \\
 1.5		& -1.23611 (-0.64618) & -0.58812 (0.00181) & -0.56588 (0.02404) & -0.58993 \\
 2.0		& -1.18750 (-0.75000) & -0.43750 (0.00000) & -0.42441 (0.01309) & -0.43750 \\
 2.5		& -1.16500 (-0.81725) & -0.34713 (0.00062) & -0.33953 (0.00821) & -0.34775 \\
 3.0		& -1.15278 (-0.86420) & -0.28688 (0.00170) & -0.28294 (0.00563) & -0.28857 \\
 3.5		& -1.14541 (-0.89879) & -0.24384 (0.00278) & -0.24252 (0.00410) & -0.24662 \\
 4.0		& -1.14062 (-0.92530) & -0.21157 (0.00376) & -0.21221 (0.00312) & -0.21533 \\
 4.5		& -1.13735 (-0.94627) & -0.18646 (0.00461) & -0.18863 (0.00245) & -0.19108 \\
 5.0		& -1.13500 (-0.96326) & -0.16638 (0.00536) & -0.16977 (0.00198) & -0.17174
\end{tabular}
\end{ruledtabular}
\end{table*}

{\em Frenkel regime.---}In the Frenkel (small-$\epsilon$) regime ($\Psi \equiv \Psi_\text{F}$ and $E \equiv E_\text{F}$), the Coulomb interaction is dominant and the electron and hole form a tightly bound pair ($\omega \approx 0$) \cite{Kayanuma92, Seidl07}. Assuming $\cot \omega \approx (2-2\cos\omega)^{-1/2} \approx \omega^{-1}$, we find
\begin{equation}
	\Psi_\text{F}^{\prime\prime}(\omega) 
	+\frac{2}{\omega}\Psi_\text{F}^{\prime}(\omega) 
	+ \left( \frac{1}{\epsilon\,\omega}+E_\text{F}\right)\,\Psi_\text{F}(\omega)=0.
\end{equation}
The above equation is a hydrogenic-like Schr\"odinger equation, and yields, for the $n$th bound state, the following zeroth-order eigenfunction and eigenvalue:
\begin{align}
	\Psi_\text{F}^{(n,0)}(\omega) & \propto L_{n}^{1}\left(\frac{\omega}{n\,\epsilon}\right) \exp\left(-\frac{\omega}{2\,n\,\epsilon}\right),
	\\
	E_\text{F}^{(n,0)} & = - \frac{1}{4\,\epsilon^2 n^2},
\end{align}
where $n \in \mathbb{N}^*$ and $L_{n}^{m}(x)$ is a generalized Laguerre polynomial \cite{NISTbook}. The lowest state ($n=1$) is a bound state, and the zeroth-order wave function $\Psi_\text{F}^{(1,0)}$ is an exponential function strongly peaked at $\omega=0$ associated with the zeroth-order energy $E_\text{F}^{(1,0)} = -1/(4\,\epsilon^2)$. Taking into account the first-order correction ($\cot \omega \approx \omega^{-1} - \omega/3$ and $(2-2\cos\omega)^{-1/2} \approx \omega^{-1} + \omega/24$) yields 
\begin{equation}
\label{E-F}
	E_\text{F}^{(1,1)} = -\frac{1}{4\,\epsilon^2} - \frac{9}{8},
\end{equation}
which is plotted in Fig.~\ref{fig:fig2}, and has the effect of stabilizing the lowest bound state. A similar calculation for an exciton on a 2-sphere \cite{Kayanuma92} yields $E_\text{F}^{(n,0)} = - 1/(\epsilon^2 (2n-1)^2)$. It shows that, similar to anisotropic semiconductors \cite{Shinada66}, the binding energy of the lowest state of the two-dimensional exciton is four times larger than in three dimensions. The similarity between flat and spherical geometries is actually not surprising because, in the  small-$\epsilon$ (large radius) limit, the surface of a 2- or 3-sphere is locally flat, and the tightly bound pair behaves as on a flat space. 

\begin{table}
\caption{
\label{tab:quasi}
Closed-form solutions of the lowest bound state and first and second excited states for an exciton on a 3-sphere.
}
\begin{ruledtabular}
\begin{tabular}{cccccc}
$n$	&	$m$	&	$a$			&	$S_{m}^{(n)}(u)$				&	$\epsilon$			&	$E$		\\	
\hline
1	&	0	&	$1$			&	$1$							&	$2$					&	$-7/16$	\\
2	&	1	&	$0$			&	$1-u/2$						&	$\sqrt{2/5}$			&	$5/4$	\\
2	&	1	&	$1$			&	$(1+(\sqrt{33}-15)u/24)$		&	$(\sqrt{33}-3)/6$		&	$9/16$	\\
2	&	2	&	$0$			&	$1-u/2+7u^2/132$			&	$\sqrt{2/33}$			&	$3$		\\
\end{tabular}
\end{ruledtabular}
\end{table}

{\em Wannier-Mott  regime.---}In the Wannier-Mott  (large-$\epsilon$) regime ($\Psi \equiv \Psi_\text{WM}$ and $E \equiv E_\text{WM}$), the kinetic energy is dominant, and the exciton is uniformly delocalized over the 3-sphere \cite{TEOAS09}. This regime can be studied using perturbation theory by treating the screened Coulomb interaction as a perturbation.
The zeroth-order wave function and energy for the $n$th state are
\begin{align}
	\Psi_\text{WM}^{(n,0)}(\omega) & = \frac{1}{2\pi^2} U_{n-1}\left(\cos\omega\right),
	\\
	E_\text{WM}^{(n,0)} & = (n^2-1),
\end{align}
where $U_{n}(x)$ is a Chebyshev polynomial of the second kind \cite{NISTbook}. 
In this regime, the lowest-energy state zeroth-order wave function $\Psi_\text{WM}^{(1,0)} = 1/(2 \pi^2)$ is a constant, yielding an equally distributed probability of finding the electron-hole pair at any point on the 3-sphere, and is associated with the zeroth-order energy $E_\text{WM}^{(1,0)} = 0$.

Taking into account the first-order correction gives 
\begin{equation}
\label{Psi-WM}
	\Psi_\text{WM}^{(1,1)}(\omega) = \frac{1}{2\pi^2} 
	+ \frac{13+6(\pi-\omega) \cot \omega - 3\pi \csc(\omega/2)}{9\,\pi^3\,\epsilon},
\end{equation}
which yields the third-order energy
\begin{multline}
\label{E-WM}
	E_\text{WM}^{(1,3)} = -\frac{8}{3 \pi\,\epsilon}
	+ \frac{4}{27\,\epsilon^2} \left(9 - \frac{92}{\pi^2}\right)
	\\
	- \frac{16}{243\,\epsilon^3} \left[\frac{2248}{\pi^3} - \frac{285}{\pi} + \frac{216}{\pi}  \ln2 - \frac{756}{\pi^3} \zeta(3)\right],
\end{multline}
where $\zeta$ is the Riemann zeta function \cite{NISTbook}. The higher-order corrections stabilize the lowest state, which becomes a bound state for any value of $\epsilon$. One can verify that, for $\epsilon\to\infty$, the wave function \eqref{Psi-WM} has the right electron-hole cusp (Eq.~\eqref{cusp}). Equations \eqref{E-F} and \eqref{E-WM} are reported in Fig.~\ref{fig:fig2}, and are compared to exact results obtained by \alert{diagonalization of Eq.~\eqref{H-omega} using a non-orthogonal basis set of the form $f_n(\omega) \propto \sin^n \omega/2$. .The lowest bound-state energy is the lowest eigenvalue of $\mathbf{S}^{-1/2} \cdot \mathbf{H} \cdot \mathbf{S}^{-1/2}$, where $\mathbf{S}$ and $\mathbf{H}$ are the overlap and Hamiltonian matrices, respectively. (See Ref.~ \cite{TEOAS09} for more details.)}

\begin{figure}
	\includegraphics[width=0.4\textwidth]{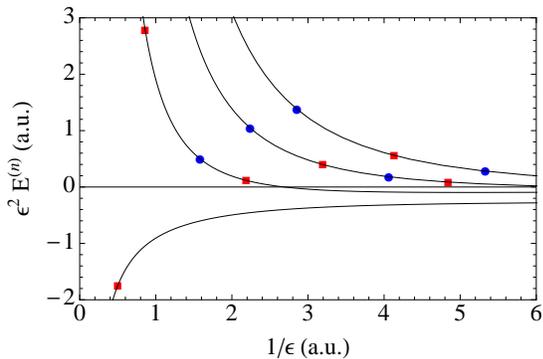}
	\caption{Exact energies of the lowest bound state and the first few excited states of an exciton on a 3-sphere with dielectric constant $\epsilon$. Closed-form solutions are shown by blue dots ($a=0$) and red squares ($a=1$).}
	\label{fig:fig3}
\end{figure}

\begin{figure}
	\includegraphics[width=0.4\textwidth]{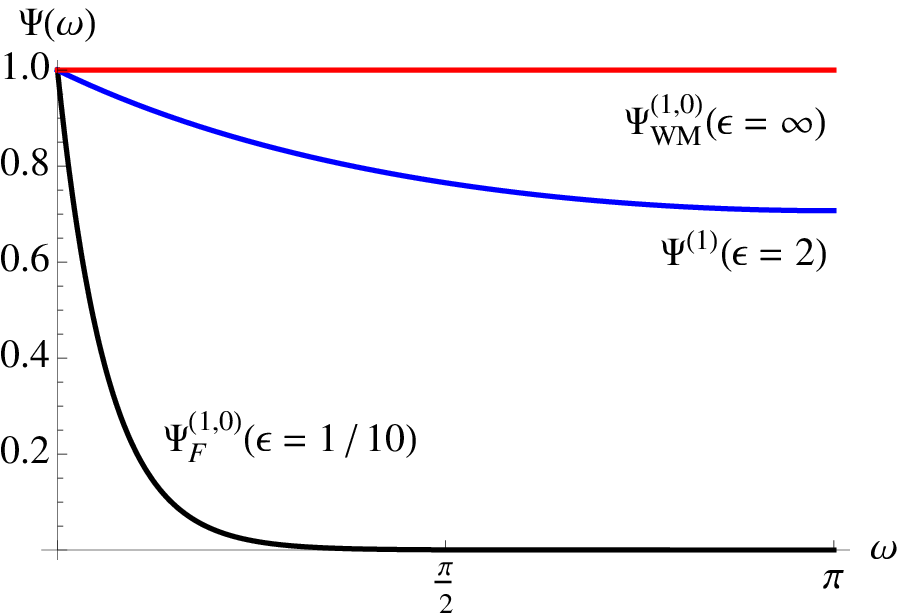}\\
	\includegraphics[width=0.4\textwidth]{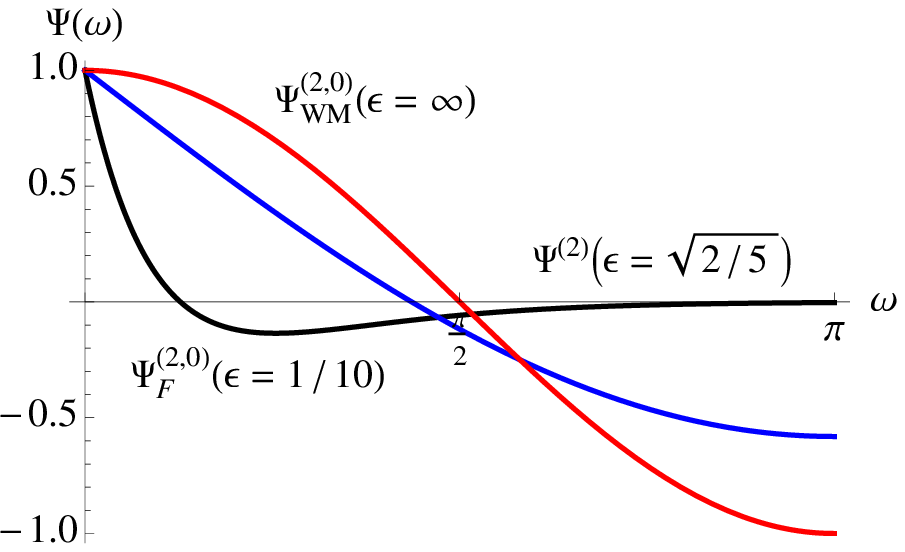}
	\caption{Frenkel (black), Wannier-Mott (red) and exact (blue) wave functions as functions of $\omega$ for the lowest bound (top) and first excited (bottom) states of an exciton on a 3-sphere.}
	\label{fig:fig4}
\end{figure}

{\em Exact closed-form solutions.---}We now turn our attention to the exact closed-form solutions which can be obtained for particular values of the dielectric constant $\epsilon$. In terms of $r=u/R$, Eq.~\eqref{H-omega} is 
\begin{equation}
\label{H-r}
	\left(\frac{r^2}{4} - 1 \right) \Psi^{\prime\prime}(r) 
	+ \left(\frac{5\,r}{4} - \frac{2}{r} \right)\Psi^{\prime}(r) 
	= \left( \frac{1}{\epsilon\,r}+E\right)\,\Psi(r).
\end{equation}
The general solution of \eqref{H-r} is \cite{NISTbook, Ronveaux} 
\begin{equation}
\label{Psi-series}
	\Psi(r) = \left(1+\frac{r}{2}\right)^{-a/2} S(r),
\end{equation}
where $a=0$ or $1$, and $S(r)$ is a regular power series i.e. $S(r) = \sum_{k=0}^{\infty} s_{k}\,r^k$. Substituting the previous series into \eqref{H-r} yields a three-term recurrence relation for the coefficients $s_k$'s. To get closed-form solutions, we assume that the series $S(r)$ terminates at a rank $m$, such as $S_{m}^{(n)}(r) = \sum_{k=0}^{m} s_{k}\,r^k$. This does happen if, and only if $s_{m+1} = s_{m+2} = 0$.  The exact energy $E$ and dielectric constant $\epsilon$ are simply given by the roots of the polynomial equations $s_{m+1} = 0$ and $s_{m+2} = 0$. We refer the reader to Ref.~\onlinecite{QR12} for more details.

This produces two families of solutions characterized by the integer $a$.  \alert{Each family contains an infinite number of solutions, associated with distinct values of $\epsilon$.} Both bound and unbound state wave functions can be obtained, and they are easily characterized by the number of nodes ($n-1$) between $r=0$ and $2$. The first few closed-form solutions are gathered in Table \ref{tab:quasi} and represented in Fig.~\ref{fig:fig3}. One can note that the lowest-energy state is associated with a negative energy for any value of the dielectric constant. This is not the case for the higher-energy states, which become bounded when the dielectric constant is large enough to screen the electron-hole attraction.

Without a doubt, the most interesting closed-form solution is
\begin{align}
\label{exact-lowest}
	\Psi^{(1)}(r) =\left(1+\frac{r}{2}\right)^{-1/2},
\end{align}
for $\epsilon = 2$ and $E^{(1)} = -7/16$. This is the unique exact wave function for the lowest bound state \footnote{We note that one cannot obtain bound state wave functions for an exciton on a 2-sphere because the value of $a$ is restricted to zero due to the dimensionality of the system.} (see Fig.~\ref{fig:fig3}). 

Another interesting wave function due to its simplicity is 
\begin{equation}
	\Psi^{(2)}(r) =1-\sqrt{\frac{5}{8}}r,
\end{equation} 
which is exact for $\epsilon = \sqrt{2/5}$, and yields $E^{(2)} = 5/4$. The three types of wave functions (Frenkel, Wannier-Mott and exact) are plotted in Fig.~\ref{fig:fig4} for the lowest bound state and the first excited state.

{\em Intermediate regime.---}We introduce a regime which is distinct from the Frenkel and Wannier-Mott asymptotic regimes. This unusual intermediate case can be experimentally observed in fluid xenon for example. \cite{Laporte80} In this intermediate regime ($\Psi \equiv \Psi_\text{int}$ and $E \equiv E_\text{int}$), the exact solution \eqref{exact-lowest} is used as a zeroth-order wave function for perturbation theory, where the zeroth-order Hamiltonian is given by Eq.~\eqref{H-r} in which $\epsilon=2$, and the perturbation operator is $(\epsilon-2)/(2\,\epsilon\,r)$. This yields, for the lowest bound state, the first-order energy
\begin{equation}
	E_\text{int}^{(1,1)} = -\frac{7}{16} + \frac{3(4-\pi)(\epsilon-2)}{4(3\pi-8)\,\epsilon} + O\left[\left(\frac{\epsilon-2}{\epsilon}\right)^2\right],
\end{equation}
which is plotted in Fig.~\eqref{fig:fig2}, and gives good agreement with the exact results for a wide range of $\epsilon$ values (Table \ref{tab:values}). Compared to the Frenkel and Wannier-Mott energy expansions truncated at the same order, the intermediate regime yields results closer to the exact values for $1 \lesssim \epsilon \lesssim 3.5$.

{\em Conclusion.---}In this Letter, we have shown that the model consisting of an exciton located on a surface of a 3-sphere is a useful model to study excitons for any value of the dielectric constant $\epsilon$. This model allows a smooth connection between the Frenkel and Wannier-Mott excitons, and we have shown that one can determine the exact closed-form expression of the exact wave function for particular values of $\epsilon$.

We thank Peter Gill, Terry Frankcombe and Joshua Hollett for useful discussions and the Australian Research Council (Grants DP0984806, DP1094170 and DP120104740) for funding.

\end{document}